\documentclass[aps,twocolumn,prb]{revtex4}
\usepackage{graphicx}
\usepackage{dcolumn}
\usepackage{amsmath}

\begin{document}

\title{Electric-field control of spin waves at room temperature in multiferroic BiFeO$_3$}

\author{P. Rovillain, M. Cazayous, Y.
  Gallais, A. Sacuto, M. A. M\'easson}
\affiliation {Laboratoire Mat\'eriaux et Ph\'enom\`enes Quantiques (UMR 7162
  CNRS), Universit\'e Paris Diderot-Paris 7, 75205 Paris cedex 13,
  France}
\author{R. de Sousa}
\affiliation {Department of Physics and Astronomy, University of Victoria,Victoria, B.C., Canada, V8W 3P6}
\author{D. Colson, A. Forget}
\affiliation {Service de Physique de l'Etat Condens\'e, DSM/DRECAM/SPEC, CEA Saclay, IRAMIS, SPEC (CNRS URA 2464), F-91191 Gif sur Yvette, France}
\author{M. Bibes \& A. Barth\'el\'emy}
\affiliation {Unit\'e Mixte de Physique CNRS/Thales, 1 av. A. Fresnel, Campus de l'Ecole Polytechnique, 
F-91767 Palaiseau, France et Universit\'e Paris-Sud, 91405  Orsay, France.}

\date{\today}

%\pacs{77.80.Bh, 75.50.Ee, 75.25.+z, 78.30.Hv}
\begin{abstract}
To face the challenges lying beyond current CMOS-based technology, new paradigms for information processing are required. {\it Magnonics}\cite{Kruglyak} proposes to use spin waves to carry and process information, in analogy with photonics that relies on light waves, with several advantageous features such as potential operation in the THz range and excellent coupling to spintronics\cite{Kajiwara}. Several magnonic analog and digital logic devices\cite{Khitun1} have been proposed, and some demonstrated\cite{Schneider}. Just as for spintronics, a key issue for magnonics is the large power required to control/write information (conventionally achieved through magnetic fields applied by strip lines, or by spin transfer from large spin-polarized currents). Here we show that in BiFeO$_3$, a room-temperature magnetoelectric material\cite{Scott}, the spin wave frequency ($>$600 GHz) can be tuned electrically by over 30\%, in a non-volatile way and with virtually no power dissipation. Theoretical calculations indicate that this effect originates from  a linear magnetoelectric effect related to spin-orbit coupling induced by the applied electric field. We argue that these properties make BiFeO$_3$ a promising medium for spin wave generation, conversion and control in future magnonics architectures. 
\end{abstract}
\maketitle

Magnetoelectric multiferroics possess coexisting magnetic and ferroelectric phases, with cross-correlation effects between magnetic and electric degrees of freedom\cite{Eerenstein}. As such, they can potentially be used to control spin-based properties by electric fields\cite{Chu}, with very low associated power dissipation. This feature appears promising not only for spintronics, in which information is encoded by the spin-polarization of the electrical current\cite{Chappert}, but also for magnonics that use magnetic excitations (spin waves) for information processing\cite{Kruglyak}. Indeed, just as coupling between magnetic and ferroelectric order parameters exists in multiferroics, coupled spin and lattice excitations  termed electromagnons have also been demonstrated\cite{Pimenov}. Such mixed excitations exist at low temperature in multiferroic manganites\cite{Pimenov} and are suspected at room temperature in BiFeO$_3$\cite{Cazayous}. Although in the former systems, electromagnons have been shown to depend on magnetic fields\cite{Pimenov1}, their predicted sensitivity to electric fields, at the heart of their potential applicative interest, remains to be demonstrated.

\begin{figure}
\includegraphics*[width=9cm]{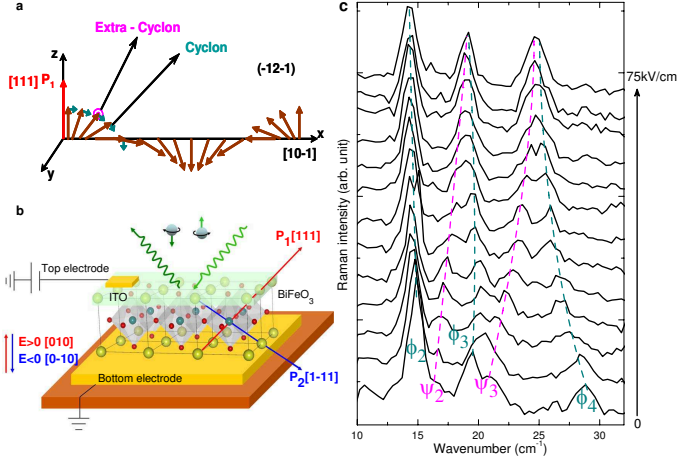}\\
  \caption{\label{Fig1} \textbf{Electrical control of spin waves.}
    \textbf{a,} Depicts the spin cycloid ground state along with its
    low energy excitations when the ferroelectric polarization
    $\bm{P}_{1}$ points along the [111] direction. The spin
    excitations correspond to in plane modes $\phi$ with an
    ellipsoidal shape elongated along y (cyclons) and out-of-plane
    modes $\psi$ (extra-cyclons) as ellipse elongated along the
    tangent vector which belongs to the xz plane.  \textbf{b,}
    Schematic diagram of the experimental set up used to apply an
    external electric field $E$ to BiFeO$_3$ single crystals and to
    probe its spin excitations using Raman spectroscopy.  The
    electrodes are deposited on the (010) plane.  A positive
    (negative) electric field parallel to the [010] ([0-10]) direction
    allows the orientation of the ferroelectric moment $\bm{P_1}$
    ($\bm{P_2}$) along the [111] ([1-11]) direction.  Raman scattering
    is performed through the indium tin oxide (ITO) top electrode with
    light polarizations in the (010) plane.  \textbf{c,} Raman spectra
    showing the magnon modes $\phi_2$, $\psi_2$, $\phi_3$, $\psi_3$
    and $\phi_4$ when the applied field ranges from 0 to 75 kV/cm.
    Dashed lines are guides to the eye following the shifts in
    frequency of each mode.}
\end{figure}

In this paper, we focus on BiFeO$_3$(BFO), a special case among multiferroic materials. Indeed, BFO has a very high ferroelectric polarization reaching 100~$\mu$C/cm$^{2}$$^[$\cite{Lebeugle}$^]$ 
below T$_C$ = 1143~K and becomes an antiferromagnet below 
T$_N$ = 643~K. The spins form a cycloid structure with a wavelength of $\lambda_0=62$~nm and an associated cycloid
wavevector equal to $Q=2\pi/\lambda_0$\cite{Smolenski, Lee, Sosnowska}.  The magnetic cycloid lies in
the (-12-1) plane, formed by the ferroelectric polarization
$\bm{P_1}\parallel [111]$ and the cycloid wavevector
$\bm{Q}\parallel[10-1]$ (Fig.~\ref{Fig1}a).  An electric field can 
simultaneously induce the flop of the polarization and the cycloidal
plane\cite{Lebeugle2}. The electrical switch of antiferromagnetic-ferroelectric domains 
in BFO has been observed\cite{Zhao, Baek} and controlled in a ferromagnet-multiferroic
heterostructure\cite{Chu}.  Here we report direct electric-field
control of spin wave states in BFO at room temperature.

In BFO, inelastic light scattering reveals two species of spin wave
excitation modes. These are observed in Raman spectra as two series of
sharp peaks labelled
cyclon ($\phi_n$) and extra cyclon modes ($\psi_n$) below 70 cm$^{-1}$, the frequency of the lowest phonon
mode\cite{Cazayous, Singh, Rovillain}. 
The cyclon $\phi_n$ and extra cyclon
$\psi_n$ modes correspond respectively to the spin oscillations in and
out of the cycloidal plane (Fig.~\ref{Fig1}a) and their Raman scattering signature are
reported in Fig.~\ref{Fig1}c.  These two species of magnons originate
from the translational symmetry breaking\cite{Sousa} of the cycloidal
ground state. The  spin wave momentum is no more conserved,
and can be increased or decreased by a multiple of the cycloid
wavevector $Q$. This leads to magnon zone folding at the Brillouin
zone center, where $\phi_n$ and $\psi_n$ correspond to magnon
wavevector equal to $\bm{k}=n\bm{Q}$.  

The experimental set-up used to apply an electric field and to detect
the optical excitations of spin waves is presented in
Fig.~\ref{Fig1}b. The device consists of bottom and top (transparent)
electrodes that enable the application of an electric field along the
$[010]$ and $[0-10]$ directions of a bulk BFO sample. The effect of the external
electric field on the spin wave modes is tracked in
Fig.~\ref{Fig1}c.  An external electric field applied along
the [010] direction induces a blue shift of the
$\psi_n$ modes ($\psi_2$ and $\psi_3$) and a red shift of the
$\phi_n$ modes ($\phi_2$, $\phi_3$, and $\phi_4$).  The promise of this
electric-field control depends on our ability to selectively tune the spin wave
states using the polarization hysteresis cycle and the switching of
the polarization vector $\bm{P}$.

\begin{figure}
\includegraphics*[width=8cm]{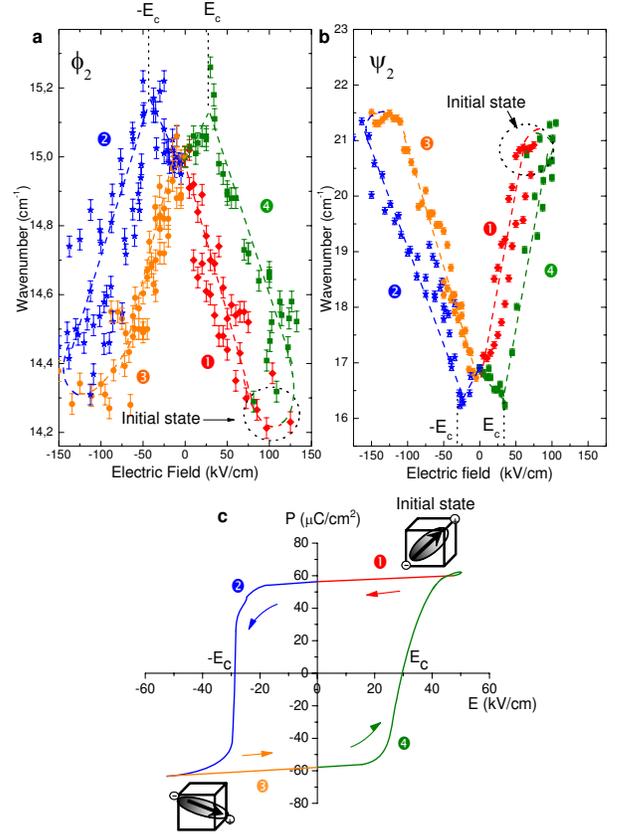}\\
  \caption{\label{Fig2} \textbf{Spin wave hysteresis loop.} Voltage
    dependence of \textbf{a,} $\phi_2$ and \textbf{b,} $\psi_2$ magnon
    modes. \textbf{c,} (P-E) hysteresis loop at room temperature. The shift of the spin wave frequencies follows the
    polarization loop along 1 (red diamond), 2 (blue star), 3 (orange
    circle) and 4 (green square).}
\end{figure}

Figures~\ref{Fig2}a,b show the frequency shift of the $\phi_2$
(cyclon) and $\psi_2$ (extra-cyclon) modes as a function of the
polarization cycle. This figure demonstrates that the spin wave modes
are directly connected to the polarization hysteresis loop shown in
Fig.~\ref{Fig2}c.  With an initial polarization state $\bm{P_1}$
$\parallel$~[111], decreasing the applied electric field ($E
\parallel$~[010]) induces an increase of the $\phi_2$ mode frequency
(red diamond, Fig.~\ref{Fig2}a) following path 1 in the polarization
cycle. Traveling along path 2 with a negative electric field
($E\parallel$~[0-10]), the frequency of the $\phi_2$ mode (blue star)
increases up to the flip of $\bm{P}$ along the [1-11] direction
($\bm{P_2}$) at critical electric field $E_c\approx -35$~kV/cm\cite{Lebeugle}. Beyond
$E_c$, $\phi_2$'s frequency decreases. The hysteresis
cycle can be completed along paths 3 and 4 leading to frequency shifts
that are symmetric to the ones along 1 and 2. 
The $\psi_2$ mode is also directly connected to the polarization loop,
but exhibits an opposite behaviour with respect to the $\phi_2$ mode
(Fig.~\ref{Fig2}b).  The extra-cyclon mode $\psi_2$ presents a much
stronger frequency shift under the effect of an electric field of
50~kV/cm, $\Delta \omega_{\psi_2}=(+2.1 \pm 0.1)$~cm$^{-1}$.
Remarkably, the $\psi_2$ frequency is seen to increase by over
$5$~cm$^{-1}$ at maximum $E$ field, corresponding to a 30\% shift of its natural frequency. 
This experiment clearly demonstrates the coupling
between the spin waves and the ferroelectric character of the material
and that a spin wave frequency hysteresis loop can be created and controlled at
room temperature using the ferroelectric hysteresis loop.

There exists other methods for electric-field control of spin waves.
Rado \textit{et al.} detected a shift of the order of $3\times
10^{-5}$~cm$^{-1}$ (relative shift of 0.01\%) in lithium ferrite under the application
of similar voltages\cite{Rado}. Recently, Vlaminck and Bailleul
demonstrated a doppler shift of the order of $10^{-3}$~cm$^{-1}$
(0.3\%) using electric currents\cite{Vlaminck}. 
The present experiment shows that the
coupling between magnetic and ferroelectric degrees of freedom
provides a much more efficient method for electric-field control of
spin waves, with frequency shifts over $5$~cm$^{-1}$ (30\%),
several orders of magnitude larger than with previous methods.

The strong spin wave frequency shifts that we observe may be interpreted as resulting from a combination of magnetoelectric interactions.
The component of $\bm{E}$ along $\bm{P}$ (denoted $E_z$) cannot play any role because of the competition with the internal electric
field parallel to $\hat{\bm{z}}$ produced by $\bm{P}$.  Hence we only
consider interactions involving $\bm{E}_\perp=(E_x, E_y)$.  
The Dzyaloshinskii-Moriya (DM) interaction that couples $\bm{E}$ to $\bm{M}\times\bm{L}$
is allowed by the R3c crystal symmetry of BFO\cite{Sousa2}, and also induces a transition from the cycloidal to a homogeneous spin state
at large $E_\perp$ \cite{Sparavigna}. 
The DM interaction produces spin wave shifts that are independent of $E$ at low electric field, at odds with the data,
and that scale as $E_{\perp}^{2}$ at larger field\cite{desousaAPL08}. 
 Another class of magnetoelectric
interactions, the so called flexoelectric
interactions\cite{Sparavigna,Mills} couple $\bm{E}$ to gradients of
$\bm{L}$. For example, the flexoelectric interaction in a homogeneous magnet leads to an
instability towards a spiral state and to spin wave shifts linear in
$\bm{E}_{\perp}$\cite{Mills}. However, in our case of bulk BFO the
magnetic order is a cycloid that is itself induced by terms that couple $\bm{P}$ to gradients of
$\bm{L}$ and have the same form as the flexoelectric interaction\cite{Sparavigna,Zvezdin09}.
Therefore, there is a competition between these different couplings that leads to a renormalization
of the cycloid wavevector. Since both
$\omega_{\phi}$ and $\omega_{\psi}$ are directly proportional to the cycloid wavevector,
a large $E_{\perp}$ will increase all spin wave frequencies by the
same amount. Thus, the DM and the flexoelectric interactions are not able to reproduce the experimental data at lower field but might 
 play a role at larger field.

\begin{figure}
\includegraphics*[width=8cm]{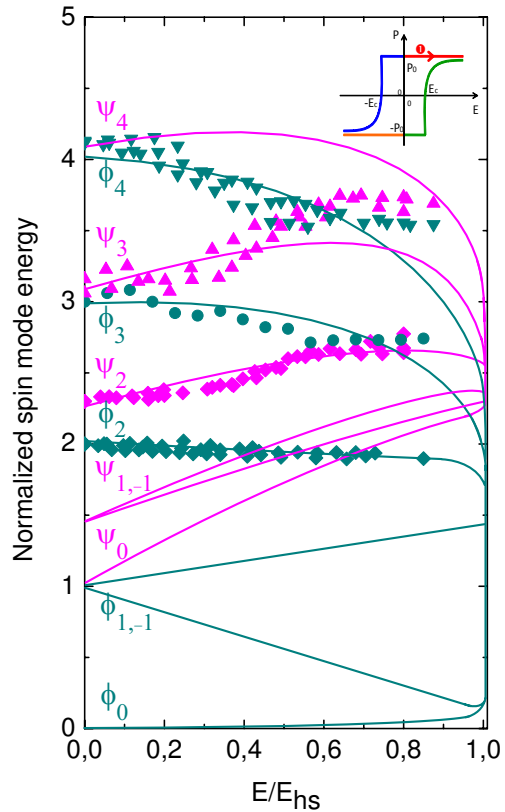}\\
  \caption{\label{Fig3} \textbf{Electric field dependence of the
      measured and calculated spin wave frequencies.}  When the
    applied electric field is zero, the energy of the $\phi_n$ and
    $\psi_n$ modes are $E_{\phi_n} = n v_0 Q$ and $E_{\psi_n} =
    \sqrt{1+n^2} v_0 Q$ respectively, with index $n$ labelling the
    modes, $v_0= 1.4 \times 10^{6}$~cm/s the fundamental cyclon spin
    wave velocity, and $Q$ the cycloid wavevector\cite{Cazayous}.  The figure shows
    the spin wave mode energies normalized to $v_0 Q$ as a function of
    the ratio of the applied electric field $E$ divided by the
    electric field $E_{hs}$ required to destroy the cycloid.  When
    $E>E_{hs}$ the spin ground state becomes homogeneous in space.
    The closed symbols are the experimental data obtained following
    path 1 (insert) in the polarization cycle using $E\parallel$~[010]
    from 0 to 125 kV/cm : $\phi_2$ (circle), $\phi_3$ (diamond),
    $\phi_4$ (down triangle), $\psi_2$ (star), $\psi_3$ (up triangle).
    ($\bm{P}$ is parallel to the [111] direction and no polarization
    flop occurs.  Solid lines are explicit numerical calculations
    obtained from a dynamical Ginzburg-Landau theory  
     based on the linear magnetoelectric interaction $F_2$.}
\end{figure}

Here, we have found that the spin wave frequency shifts can be interpreted using the Landau free energy model
based on an additional kind of magnetoelectric interactions induced by 
the external electric field.
Indeed, the R3c
symmetry of BFO allows the presence of two linear magnetoelectric
interactions that couple the N\'{e}el order parameter $\bm{L}$
directly with $\bm{E}_{\perp}$,

Here, we have found that the spin wave frequency shifts can be interpreted using the Landau free energy model
based on an additional kind of magnetoelectric interactions induced by 
the external electric field.
Indeed, the R3c
symmetry of BFO allows the presence of two linear magnetoelectric
interactions that couple the N\'{e}el order parameter $\bm{L}$
directly with $\bm{E}_{\perp}$,
\begin{eqnarray}
  F_{1}&=& \frac{1}{2}\kappa_1 \bm{E}_{\perp}\cdot \bm{L}_\perp L_z, \label{f1}\\
  F_{2}&=& \frac{1}{2}\kappa_2 \bm{E}_{\perp}\cdot 
  \left[\left(L_{y}^{2}-L_{x}^{2}\right)\hat{\bm{x}}+2L_{x}L_{y}\hat{\bm{y}}\right],\label{f2}
\end{eqnarray}
with $\kappa_1$ and $\kappa_2$ phenomenological coupling constants.
Physically, $F_1$ and $F_2$ model additional magnetic anisotropy
energies induced by $E_\perp$. Hence their microscopic origin is related
to spin-orbit coupling.  
The first interaction
($F_1$) produces a red shift for both $\psi_n$ and $\phi_n$ modes when
$E_\perp$ is increased.  Thus, $F_1$ cannot explain the experimental
observations at low $E_{\perp}$ (Fig.~\ref{Fig2}a,b). On the other
hand, the interaction $F_{2}$ induces a blue shift for all $\psi_n$
modes, and a red shift for $\phi_n$ modes with $n\geq 2$ as found experimentally, see Fig.~\ref{Fig3}.
This interaction drives
the system into a homogeneous state when $E_\perp$ becomes greater
than a homogeneous critical field $E_{hs}\propto 1/\kappa_2$. The
calculated curves show that cyclon and extra cyclon mode
frequencies converge to two fixed points at the critical field
$E_{hs}$. 
A comparison between theory and experiment (Fig.~\ref{Fig3})
gives an estimate of $E_{hs}\approx 160$~kV/cm (above the maximum electric field of $125$~kV/cm that could be
applied to our sample without damage).  Calculations are in good
agreement with the experimental data in the low field region (see
Fig.~\ref{Fig3}), but a clear deviation occurs at high fields.
Interestingly, cyclon and extra cyclon modes seem to merge for $E>0.5 E_{hs}$. This effect and the difference between theory and experiment at high fields might be due to other magnetoelectric interactions such as the DM and the flexoelectric interactions discussed above.

The demonstration of electrical control of spin wave states represents
a significant step towards making new spin wave-based technologies.
The cyclon spin wave dispersion can be written as
$\omega_{\phi}=v_{\phi} k$, and the extra-cyclon dispersion as
$\omega_{\psi}=v_{\psi} \sqrt{k^2+Q^2}$. For $E=0$, the spin wave
velocities are equal, $v_{\phi}=v_{\psi}=v_0=1.4\times 10^{6}$~cm/s
\cite{Cazayous}. However, under an electric field of $100$~kV/cm the
$\phi$ and $\psi$ spin wave velocities are tuned by $\Delta v_{\phi}=
-6 \times 10^{4}$~cm/s, and $\Delta v_{\psi}= +4\times 10^{5}$~cm/s
respectively.  While the speed of electron spin precession in a
magnetic field is generally fixed by intrinsic properties of the
materials considered, our work shows that in a multiferroic the spin
wave speed can be continuously adjusted and in a different way
depending on the spin propagating mode by the applied electric fields.

We have demonstrated that the frequency of spin waves can been tuned electrically by over 30\% at room temperature in multiferroic BiFeO$_3$. This electric-field dependence mimics that of the ferroelectric polarization, providing a handle for the non-volatile, low-power control of spin waves. Thus, aside from its potential for spintronics\cite{Bea} and photonics\cite{Choi}, BiFeO$_3$ emerges as an exciting platform for testing novel magnonic device concepts, which further confirms it as a key multifunctional material for beyond-CMOS technology.

\section*{Methode}

\subsection*{Experimental method}

  BiFeO$_3$ single crystals were grown in air using
  Bi$_2$O$_3$-Fe$_2$O$_3$ flux in an alumina crucible as detailed in ref \cite{Lebeugle}.
  and present a single ferroelectric and
  antiferromagnetic domain state. 
  Polarized optical microscope images of the samples show that the crystal consists of one single ferroelectric domain\cite{Lebeugle}.
Neutron measurements on the same samples have shown the presence of one antiferromagnetic domain\cite{Lebeugle2}.
  A conducting Indium tin oxide (ITO) layer was grown by pulsed laser deposition on the top of
  the (010) sample surface to apply a uniform electric
  field. ITO has no Raman signal in the frequency
  range of interest.  
  All the measurements were performed in vacuum at room temperature on several samples. 
  
  Raman scattering was performed on the (010) sample surface in
  backscattering geometry using the $647.1$~nm laser line. Raman scattering is collected by a
  triple spectrometer Jobin Yvon T64000 equipped with a CCD. The spot
  size is about $100$~$\mu$m$^2$ and the penetration depth is less
  than $10^{-5}$~cm. Reproducible Raman measurements have been performed on 
  several points on the sample surfaces. 
  
  The $\phi_n$ and $\psi_n$ modes ($n$ index labels the modes from their lowest to highest energy) were selectively observed using parallel and crossed polarizations in the (010) plane, respectively (see Supplementary Information). The polarizations are defined with respect to the projection of the cycloid wavevector $Q$ $\parallel$ [10-1] in the (010) plane (the direction of the wavevector $Q$ is not directly accessible in the (010) plane). 
In order to show simultaneously several $\phi_n$ and $\psi_n$ modes, Fig.~\ref{Fig1}b presents measurements without specific polarization.

\subsection*{Theoretical method}
 We consider a total free
  energy given by $F=F_0+F_1+F_2$, with $F_0$ the usual model free
  energy of multiferroic BFO \cite{Sparavigna,Sousa},\\
\begin{eqnarray}
F_0 = \frac{A}{2}L^{2}+\frac{r}{2}M^{2}+\frac{G}{4}L^{4}+ 
 \frac{c}{2}\sum_{i=x,y,z} 
(\nabla L_i)^{2}\nonumber \\
-\alpha_P\bm{P}\cdot[\bm{L}(\nabla
\cdot \bm{L})
+\bm{L}\times (\nabla \times \bm{L})],
\label{f0}
\end{eqnarray}\\
and $F_1$, $F_2$ given by Eqs.~(\ref{f1})~and~(\ref{f2}) respectively
[we omit the purely ferroelectric contributions to Eq.~(\ref{f0})
because they play no role in the discussion below].  Here $A$, $r$,
$G$, $c$, and $\alpha_P$ are phenomenological constants.  The ground
state is determined by the condition $\frac{\delta F}{\delta
  \bm{L}}=0$; when $E_{\perp}=0$ the ground state is a harmonic
cycloid
$\bm{L}_{0}=L_0[\sin{(Qx)}\hat{\bm{x}}+\cos{(Qx)}\hat{\bm{z}}]$ with
wavevector $Q=\frac{\alpha P}{c}$ and amplitude
$L_{0}^{2}=(-A+cQ^{2})/G$.  The effect of $E_{\perp}$ is to reduce the
ground state wavevector $Q$ and to add additional anharmonic
contributions to $\bm{L}_0$ at odd multiples of $Q$. When $E_{\perp}$
is equal to a critical field $E_{hs}=\frac{(\pi\alpha_P
  P)^{2}}{4c}\frac{1}{\kappa_2}$, the cycloid wavevector $Q$ becomes
equal to zero, signaling a transition to a homogeneous state.

The spin wave excitation frequencies are determined from the
Landau-Lifshitz equations of motion. After combining the equations of motion for
the order parameters $\bm{L}$ and $\bm{M}$ we get
\begin{equation}
\frac{\partial^{2}\bm{L}}{\partial t^{2}}=-r(\gamma L_0)^{2}
\left[\frac{\delta F}{\delta \bm{L}}-\left(\hat{\bm{L}}_{0}\cdot 
\frac{\delta F}{\delta \bm{L}} \hat{\bm{L}}_0\right)\right],
\label{ll}
\end{equation}
where $\gamma$ is the gyromagnetic ratio. We find wave-like solutions
by plugging in $\bm{L}=\bm{L}_0 +(\delta \bm{L})\textrm{e}^{i\omega
  t}$ in Eq.~(\ref{ll}), with
$(\delta\bm{L})=\phi(x)\bm{\hat{D}}+\psi(x) \bm{\hat{y}}$.  The
unitary vector $\bm{\hat{D}}$ is tangential to $\bm{L}_0$, i.e. it is
orthogonal to $\bm{L}_0$ but lies within the cycloid plane. 
Plugging $\phi(x)=\sum_n
\phi_n \textrm{e}^{inQx}$, and $\psi(x)=\sum_n \psi_n
\textrm{e}^{inQx}$ into Eq.~(\ref{ll}) leads to a
matrix equation whose eigenvalues give the spin wave excitations at $k=0$.

We want to underline that the free energy model for BFO Eq.~(\ref{f0}) does not contain any magnetic anisotropy. Nevertheless, the present experiment shows that an external E field induces magnetic anisotropy linear in E, see Eq.~(\ref{f2}). In other words, the electric field activates a latent anisotropy in the material, leading to a large effect in the case of Eq.~(\ref{f2}).

\section*{Acknowledgements} The authors would like to thank R. Lobo and P.
  Monod for fruitful discussions and E. Jacquet for technical assistance. DC, MB and AB would like to acknowledge support from the French Agence Nationale pour la Recherche, contract MELOIC (ANR-08-P196-36). RdS would like to acknowledge support from the Natural Sciences and Engineering Research Council of Canada.
\section*{Correspondence} Correspondence and requests for materials
should be addressed to M. Cazayous~(email: maximilien.cazayous@univ-paris-diderot.fr).

\end{document}